\documentclass[twoside,english,preprint,sort&compress]{revtex4-1}
\usepackage[T1]{fontenc}
\usepackage[latin9]{inputenc}
\usepackage[a4paper]{geometry}
\usepackage{textcomp}
\usepackage{amstext}
\usepackage{graphicx}

\makeatletter 

\@ifundefined{date}{}{\date{}}

\makeatother

\usepackage{babel}
\begin{document}

\title{Development of a strontium optical lattice clock for the SOC mission
on the ISS}

\author{K. Bongs, Y. Singh, L. Smith, W. He, O. Kock, D. \'{S}wierad, J.
Hughes,\textsuperscript{(1)}\\
 S. Schiller, S. Alighanbari, S. Origlia,\textsuperscript{(2)}\\
S. Vogt, U. Sterr, Ch. Lisdat,\textsuperscript{(3)}\\
R. Le Targat, J. Lodewyck, D. Holleville, B. Venon, S. Bize,\textsuperscript{(4)}\\
G.~P. Barwood, P. Gill, I.~R. Hill, Y.~B. Ovchinnikov,\textsuperscript{(5)}\\
 N. Poli, G. M. Tino,\textsuperscript{(6)}\\
 J. Stuhler, W. Kaenders,\textsuperscript{(7)}\textit{ and the SOC2
team}}

\address{(1) University of Birmingham (UoB), Birmingham B15 2TT, United Kingdom }

\address{(2) Institut fur Experimentalphysik, Heinrich-Heine-Universitat Dusseldorf
(HHU), 40225 Duesseldorf, Germany}

\address{(3) Physikalisch-Technische Bundesanstalt (PTB), 38116~Braunschweig,
Germany}

\address{(4) SYRTE, Observatoire de Paris, 75014 Paris, France}

\address{(5) National Physical Laboratory (NPL), Teddington, United Kingdom}

\address{(6) Universita di Firenze (UNIFI) and LENS, Firenze, Italy}

\address{(7) TOPTICA Photonics AG, 82166 Graefelfing, Germany}
\begin{abstract}
Ultra-precise optical clocks in space will allow new studies in fundamental
physics and astronomy. Within an European Space Agency (ESA) program,
the \textquotedblleft Space Optical Clocks\textquotedblright{} (SOC)
project aims to install and to operate an optical lattice clock on
the International Space Station (ISS) towards the end of this decade.
It would be a natural follow-on to the ACES mission, improving its
performance by at least one order of magnitude. The payload is planned
to include an optical lattice clock, as well as a frequency comb,
a microwave link, and an optical link for comparisons of the ISS clock
with ground clocks located in several countries and continents. Within
the EU-FP7-SPACE-2010-1 project no. 263500, during the years $2011-2015$
a compact, modular and robust strontium lattice optical clock demonstrator
has been developed. Goal performance is a fractional frequency instability
below $1\times10^{-15}\,\tau^{-1/2}$ and a fractional inaccuracy
below 5\texttimes 10\textsuperscript{-17}. Here we describe the current
status of the apparatus' development, including the laser subsystems.
Robust preparation of cold $^{88}$Sr atoms in a 2\textsuperscript{nd}-stage
magneto-optical trap (MOT) is achieved.

\textbf{}
\end{abstract}
\maketitle

\section{Introduction}

The mission ACES (Atomic Clock Ensemble in Space) \citep{hes_aces_2011}
will operate a cold-atom microwave clock on the ISS, with a planned
flight date in 2016. A follow-on mission with an improved clock was
proposed in 2005, in the framework of ESA's ISS utilization program
ELIPS. In this mission, the clock and the time/frequency links will
have a performance at least a factor of 10 better than the ACES mission.
An optical lattice clock \citep{Poli2013} is a suitable solution
for an improved clock. The payload will contain a frequency comb for
transformation of the ultra-stable optical frequency into a microwave,
in order to allow use of the microwave link technology for time and
frequency transfer between Earth and ISS. 

Since 2007, technical developments on a robust and compact Sr optical
clock are being undertaken by the Space Optical Clock (SOC) consortium
(www.soc2.eu) Important requirements for an optical clock for space
are: performance, compact size and moderate mass, radiation hardness,
fully automatic operation, low down-time, ability to withstand shocks
and vibrations during launch (when in non-operational mode), and switch
from non-operational to operational mode. These requirements led to
the development of a first-generation breadboard system which successfully
operated with $^{88}$Sr \citep{pol14}. Based on the experience gained
with that system and its subunits, the development of an advanced
breadboard system with improved specifications was initiated; its
status is reported here. Earlier developments are reported in Ref.~\citep{schiller_space_2012}

\section{The SOC mission concept}

The ISS mission SOC will implement two objectives in fundamental physics,
the measurement of the gravitational redshift in the Earth\textquoteright s
field and in the Sun\textquoteright s field. In addition, it will
be operated as a reference clock in space, combined with a high-performance
link allowing distribution of precise frequency over a large part
of Earth and allowing comparisons between distant ground clocks of
high performance, opening up the field of space-assisted relativistic
geodesy.

The precise measurement of the gravitational redshift in the fields
of two dissimilar bodies (the constitution of the atomic nuclei in
the Earth and Sun being strongly different (mostly iron vs. hydrogen))
represents a search for the existence of new fundamental fields that
induce a non-universality of the gravitational redshift effect. This
implies a strong test of Einstein\textquoteright s theory of General
Relativity as well as of the Einstein Equivalence Principle. It also
paves the way for future application of the redshift effect for high-accuracy
mapping out the gravitational potential of planets or stars. In detail,
the objectives of SOC are:
\begin{itemize}
\item Objective I\\
The measurement of the gravitational redshift of the Earth will be
performed with an accuracy improved by a factor 10 compared to the
goals of the ACES mission. The improvement factor will be limited
by the inaccuracy of the space optical clock ($1-2\times10^{-17}$)
or of the link, since a sufficient number of primary terrestrial optical
clocks having an accuracy higher than the space clock will be available
by the time of the mission \citep{Poli2013}. 
\item Objective II\\
By comparing pairs of terrestrial clocks located at a large distance
in east-west direction it is possible to perform a test of the equivalence
principle in the gravitational field of the Sun \citep{hoffmann_noon-midnight_1961,krisher_gravitational_1996}.
As any clock on the Earth is in free-fall with respect to the Sun
(neglecting tidal forces), any relative frequency shift between two
clocks caused by the Sun is expected to cancel. This is due to a cancellation
between the pure gravitational effect and the relativistic Doppler
shift occurring in a comparison between any two clocks located at
a distance. Basically, the comparisons are performed in two orientations
of the Earth: in one, the baseline between the terrestrial clocks
is perpendicular to the direction to the Sun. This frequency comparison
yields the difference in Earth\textquoteright s gravitational potential
between the two clock locations. The second orientation is when the
clocks\textquoteright{} baseline lies parallel to the direction Earth-Sun.
A measurement in this orientation contains a contribution of the Sun\textquoteright s
gravitational potential (solar redshift), but is canceled by the Doppler
shift due to the motion of the clocks along the Earth orbit. In practice,
the measurements will be performed continuously, for various orientations.
Assuming ground clocks with accuracy of $1\times10{}^{-18}$ spaced
one Earth radius away, and that a sufficiently large number of Earth
rotations and comparisons is used to reduce the inaccuracy by a factor
10 compared to a single comparison, a measurement of the combined
effect of solar redshift and Doppler shift with relative inaccuracy
of $2\times10{}^{-7}$ can be obtained. The improvement compared to
the mission ACES, which will also be capable of such a measurement,
is a factor of 10 or more, limited by the inaccuracy that distant
terrestrial clocks comparable via ACES will have achieved by the time
of its flight, and by the ground-ISS link inaccuracy.
\item Objective III\\
The Earth gravitational redshift is also the foundation for relativistic
geodesy. Terrestrial clocks and corresponding receiver systems (occupying
a volume on the order of a container or less) will eventually become
available for transportation to locations of particular geophysical
interest and can be compared to the space clock, allowing determination
of the local value of the gravitational potential. These transportable
clocks may well have reached an accuracy of $1\times10{}^{-18}$ by
the time of the SOC mission. By relying on the goal inaccuracy of
the space clock, the correctness of the gravitational redshift established
by the mission, and precise ISS orbit determination, measurements
of the local terrestrial gravitational potential at the equivalent
uncertainty level of 10~cm would be possible. However, the space
clock can also enable comparison of distant terrestrial clocks with
accuracy compatible with $1\times10{}^{-18}$ clock inaccuracy after
approximately 1~day of integration time, using an enhanced microwave
link or the optical link. Thus, the differential gravitational potential
between two terrestrial clocks may be measurable at the 1~cm uncertainty
level. Tidal effects will have to be corrected for. Note that the
relative resolution of the gravitational redshift of terrestrial clocks
corresponding to this uncertainty level (1~cm relative to a few km
maximum height difference) is on the order of several parts in $10^{6}$,
a level for which the correctness of the gravitational redshift will
have already been tested by ACES. 
\item Objective IV\\
The dissemination of ultrastable frequencies over the Earth, where
the frequencies are generated from both the space clock and a set
of ultraprecise terrestrial clocks. This objective is foreseen to
satisfy new future ground or space users. The measurement procedures
will be similar to the ones of the other objectives. Thus, the SOC
mission will contribute to link terrestrial clocks into a global network
allowing ground-to-ground or ground to space comparisons with a relative
frequency uncertainty level of $1\times10{}^{-18}$.
\end{itemize}

\section{Overview}

\begin{figure}
\includegraphics[width=0.6\columnwidth]{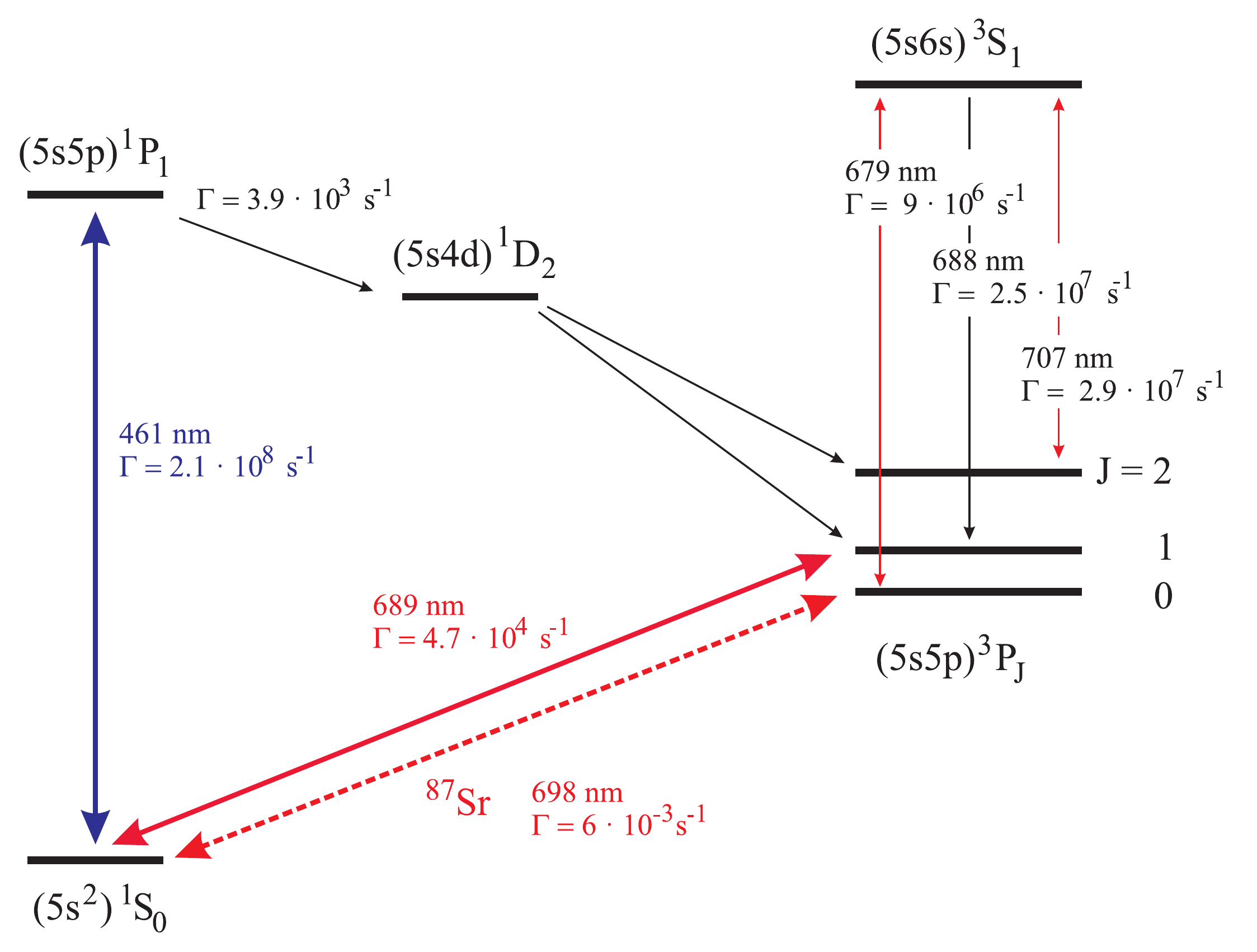}\protect\caption{Simplified energy diagram for strontium including relevant optical
transitions and spontaneous decay rates. Black lines indicate spontaneous
decay processes. \label{fig:Simplified-energy-diagram}}
\end{figure}

The transitions relevant for an optical lattice clock using strontium
are shown in Fig.\ \ref{fig:Simplified-energy-diagram}. 6 different
laser wavelengths are necessary. The apparatus described here is modular
and consists of (i) atom preparation unit, (ii) clock laser (698~nm),
(iii) cooling laser (461~nm) with distribution unit, (iv) 2\textsuperscript{nd}-stage
cooling laser (689~nm), (v) stirring laser (689~nm), (vi) repumper
lasers (679~nm, 707~nm), (vii) frequency stabilization system (FSS)
\citep{nev14}. These subunits are interconnected by optical fibers.
A computer system based on an field programmable gate array (FPGA)
controls the system. A robust wavemeter would likely be part of the
optical clock in space. However, the SOC consortium did not undertake
a corresponding development. We describe the units in turn and present
first characterization results.

\section{Subunits of the Clock}

\subsection{Atomics unit}

\begin{figure}
\includegraphics[scale=0.2]{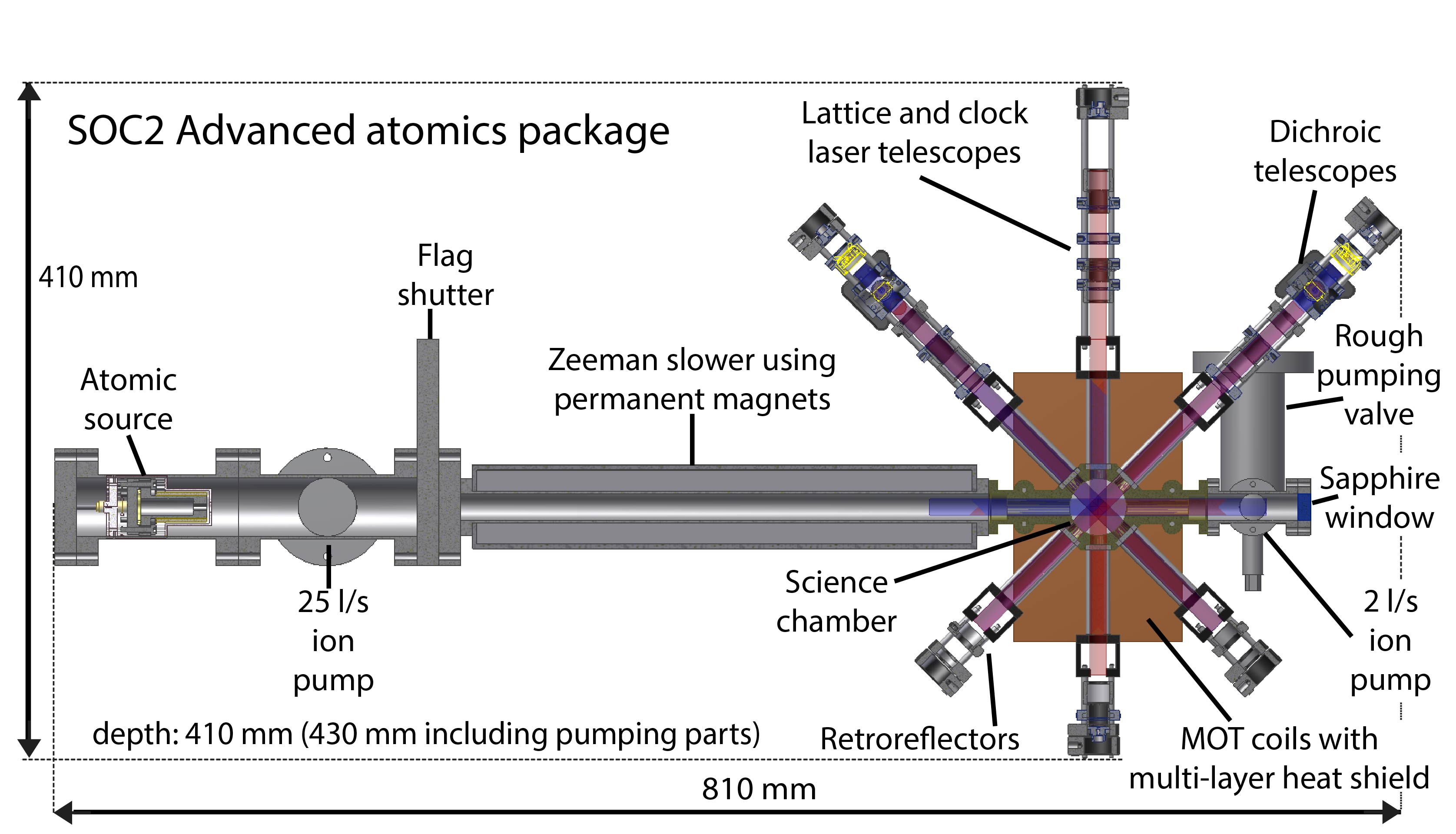}\protect\caption{\label{fig:Vacuum-assembly-for}The atomics unit. }
\end{figure}

The compact and lightweight vacuum chamber is shown in Fig.~\ref{fig:Vacuum-assembly-for}.
Including the breadboard, the overall size is 143~liter and its mass
is 50~kg. In the 3D MOT ``science'' chamber (right), an ultra-high
vacuum in the range of 10\textsuperscript{-11}~mbar is maintained.
This is achieved by employing two ion pumps (25~l/s and 2~l/s),
near to the atomic source and to the science chamber, respectively.
In order to minimize the black-body radiation (BBR) coming from the
atomic source during clock operation, an automated full-closure flag
shutter is installed. Permanent magnets are used for the Zeeman slower,
as low energy consumption is important. The Zeeman slower is described
in detail in Ref.~\citep*{2014Ovchinnikov}. 

\begin{figure}
\includegraphics[scale=0.22]{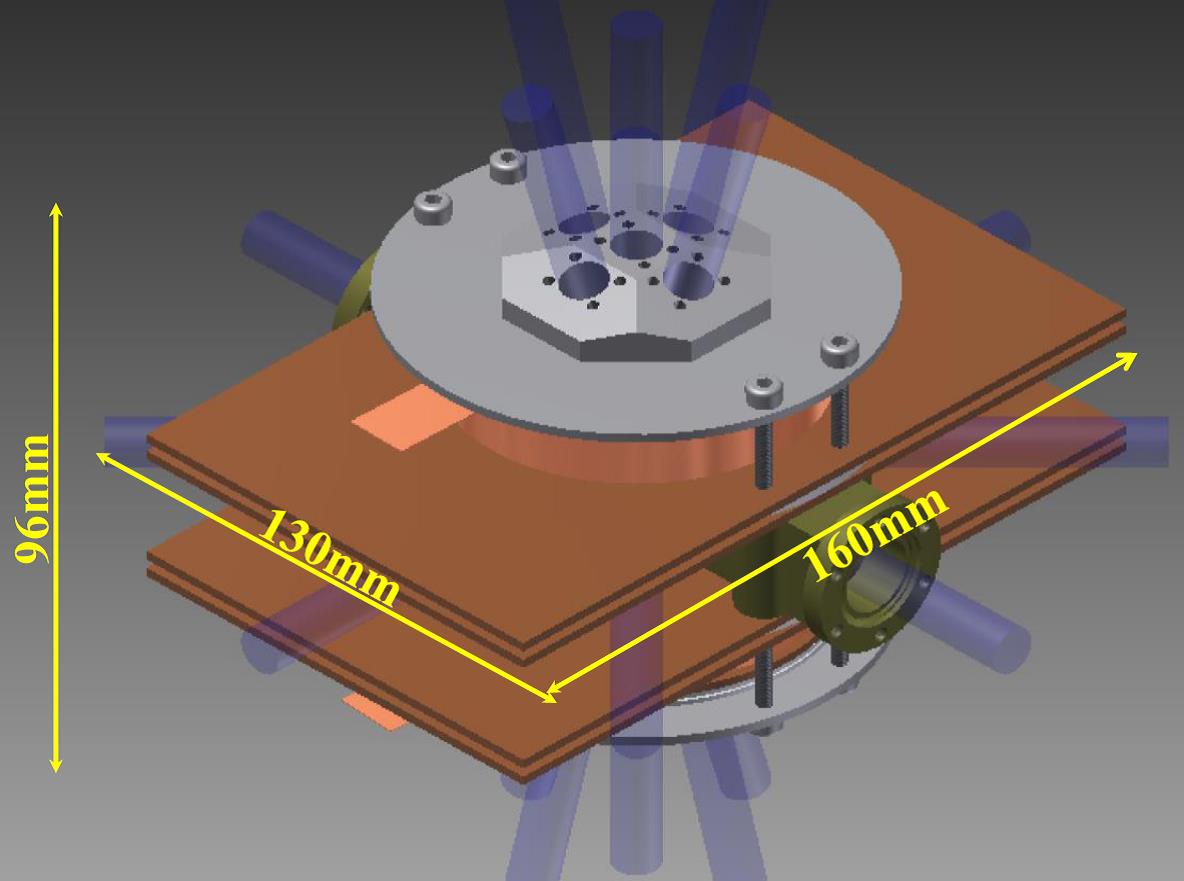}~~~~~~~\includegraphics[scale=0.22]{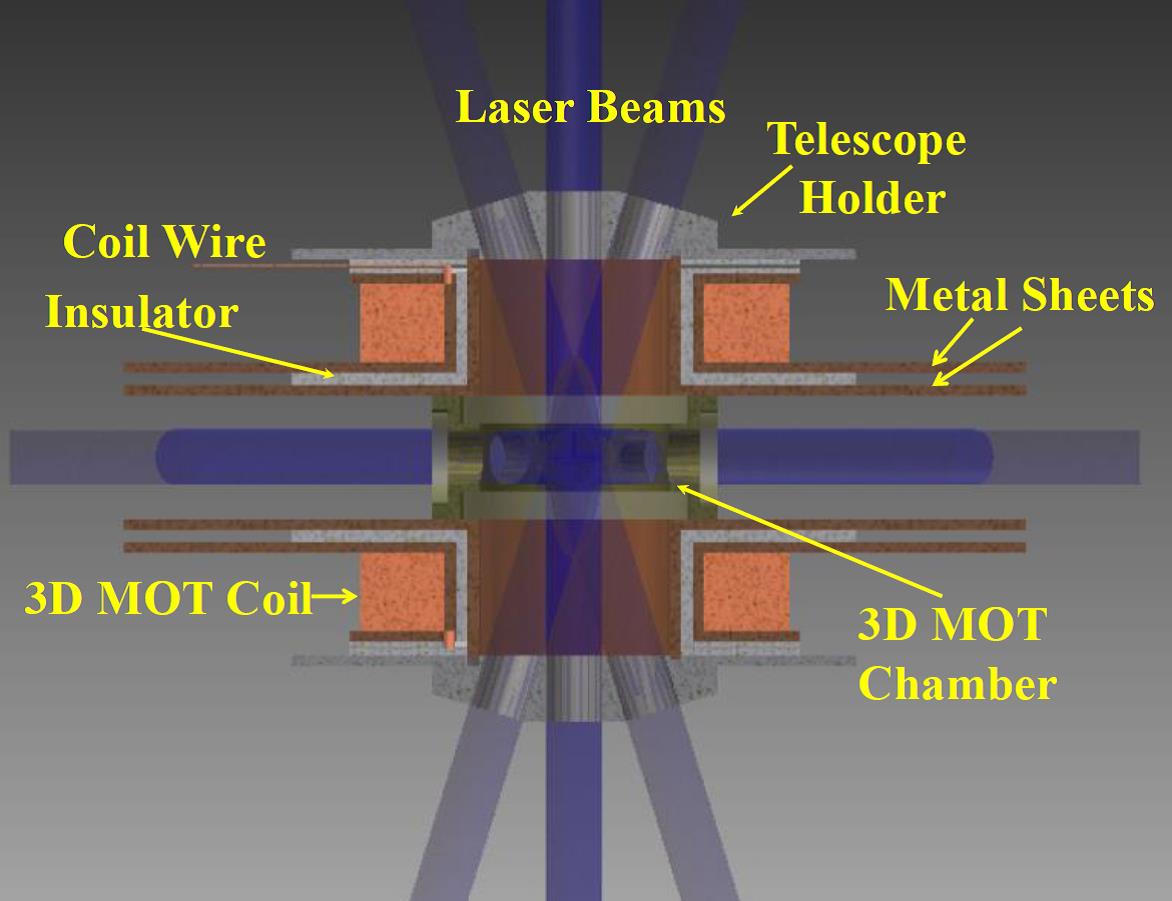}

\protect\caption{\label{fig:(Left)-3D-MOT}Left: 3D MOT chamber design. Right: cross
sectional view.}
\end{figure}

The science chamber, shown in Fig.~\ref{fig:(Left)-3D-MOT}, is very
compact compared to usual chambers for optical lattice clocks. Nevertheless,
it maintains robustness and is energy efficient. It weighs only 208~g,
has 8 viewports and a total of 17 optical accesses: six for the MOT
beams, one for the Zeeman slowing beam, two for the repump beams,
two for the 1D lattice beams/interrogation beam, two for detection
and four spare accesses for further detection methods. Six viewports
have a clear access of 10 mm diameter, and two large viewports have
a clear access of 34 mm diameter. Both of the latter contain five
accesses each of 10 mm diameter. The viewports are made from BK7 glass,
AR coated for all relevant wavelengths and sealed to the chamber using
indium. The chamber is manufactured from titanium as its thermal expansion
coefficient is similar to that of BK7 and it has favorable magnetic
properties. 

For a MOT field gradient of 4.5~mT/cm, the power dissipation per
coil is less than 6~W, yielding a total dissipation of 12~W. At
this level there is no need for active cooling of the coils. Instead,
a multi-layer heat shield for the coils has been designed and implemented.
This consists of two layers of copper separated by an insulating layer
acting in order to reduce the heating effects of the MOT coils on
the chamber (Figs.~1,~2). This heat shield could be actively temperature-controlled
in the future, in order to reduce the clock uncertainty contribution
due to BBR. 

The atomic source is a low-power oven adapted from the device described
in Ref.~\citep{sch12c}. The atomic beam deposits Sr atoms on the
viewport of the 4-way cross after the 3D MOT chamber. In order to
be able to remove the deposited strontium, a sapphire viewport was
installed, which is bakeable to high temperature for that purpose. 

\subsection{Laser system for the 1\protect\textsuperscript{st}-stage MOT}

\begin{figure}
\includegraphics[scale=0.6]{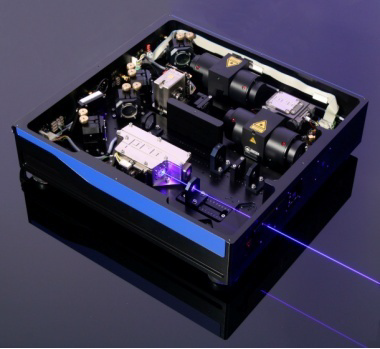}~~~~~~~~~~\includegraphics[scale=0.08]{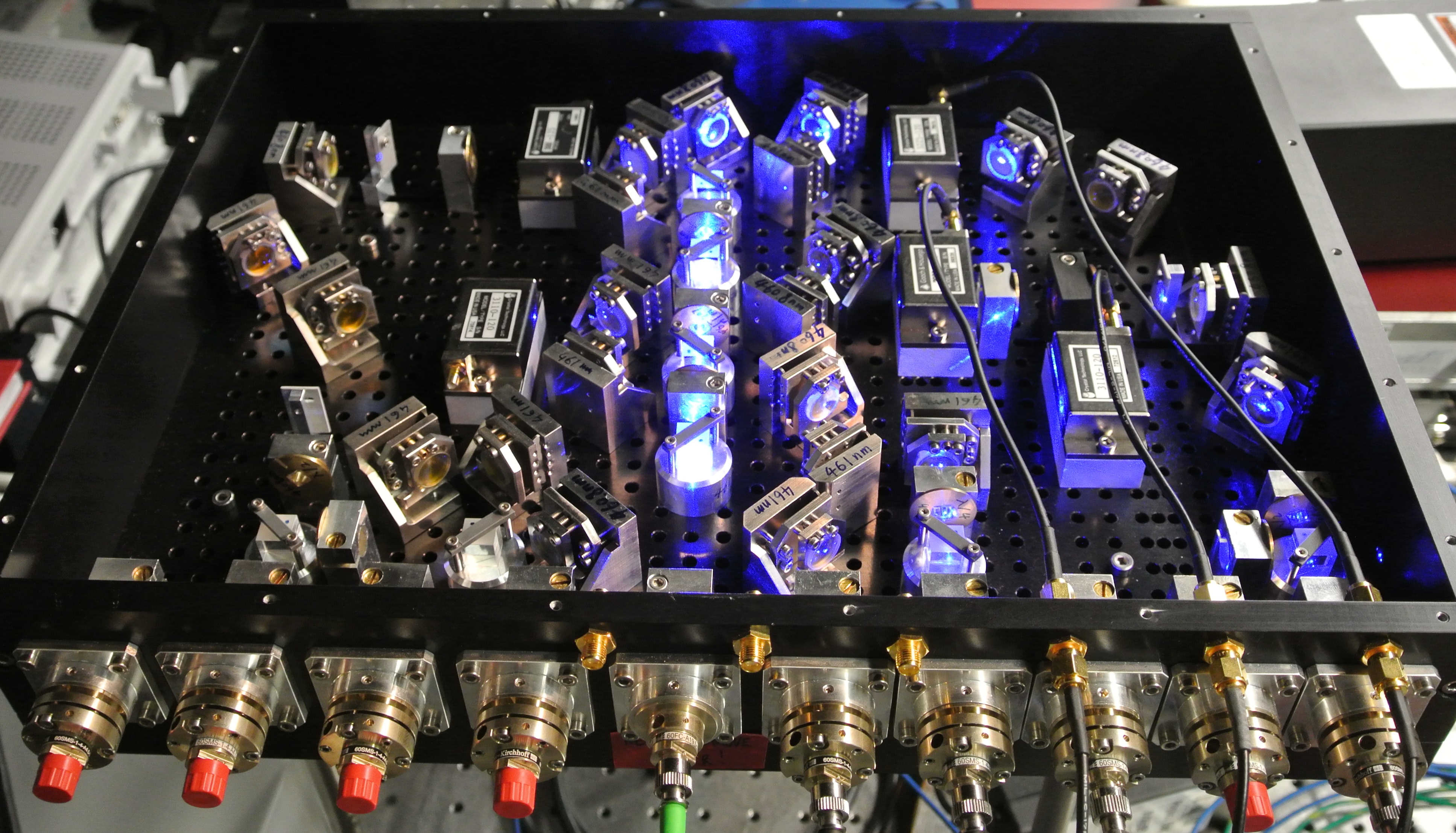}

\protect\caption{\label{fig:Left:-SHG-TA-PRO-laser}Left: SHG-TA-PRO laser system emitting
461~nm for first-stage cooling of Sr (TOPTICA). Right: 461~nm light
distribution unit. }
\end{figure}

The 461~nm laser is a TOPTICA SHG-TA-PRO system (Fig.~\ref{fig:Left:-SHG-TA-PRO-laser},
left). It contains a 922~nm laser whose output is resonantly frequency-doubled
in an enhancement cavity, emitting 400~mW. The output frequency is
stabilized to the FSS (see below). 

A compact and robust frequency distribution unit (Fig.~\ref{fig:Left:-SHG-TA-PRO-laser},
right) acts as an interface between the 461~nm laser and the atomics
unit. The unit is of size 30~cm \texttimes{} 20~cm \texttimes{}
10~cm (6~liter) and mass 5~kg and has one fiber input for the 461~nm
laser radiation and 9 fiber outputs, all of which are mounted on the
same side, for ease of access. 250~mW laser power is available in
the fiber input to the distribution unit. The fiber outputs are connected
to the atomics unit.

The module can simultaneously supply required frequency (detuning)
and intensity control for 2D \& 3D MOTs, for the Zeeman slower, for
detection and for spectroscopy that serves to lock the laser. It contains
5 acousto-optical modulators (AOMs) that can be controlled via a direct
digital synthesizer (DDS), which itself can be interfaced with a computer. 

\subsection{Lasers for the 2\protect\textsuperscript{nd}-stage MOT}

The second-stage cooling employs the $^{1}\mathrm{S}_{0}\,-\,^{3}\mathrm{P}_{1}$
transition at 689~nm with a natural linewidth of about 7.5~kHz.
In order to take full advantage of the correspondingly low Doppler
temperature, a cooling laser with sub-kHz linewidth is required. We
have observed that the spectral purity is especially important in
the frequency range extending to 100~kHz around the carrier: sidebands
or noise in this range can severely compromise the laser cooling and
thus hamper the transfer efficiency of the atoms from the 2\textsuperscript{nd}-stage
MOT into the optical lattice. The spectral purity is typically achieved
by a fast frequency lock to a highly stable optical reference resonator
in combination with sufficiently small frequency noise of the free-running
laser. Here, the FSS (Sec.~\ref{sub:FSS}) serves as such a reference.
To cover a broad range of atomic velocities when capturing atoms precooled
in the 461~nm MOT stage, during the first phase of the second-stage
cooling, the laser spectrum is broadened by modulation of the frequency
with peak-to-peak amplitude of 5MHz and modulation frequency of 30~kHz.
In our design, the modulation is introduced within the frequency
control servo loop to the reference cavity through frequency modulation
of the sideband frequency in the FSS, thus avoiding the power consumption
and complexity of a dedicated double-pass AOM on the laser breadboard.
Furthermore, the 2\textsuperscript{nd}-stage cooling of \textsuperscript{87}Sr
requires not only one laser that addresses the $\Delta F=+1$ cooling
transition but a second (so-called stirring) laser to drive the $\Delta F=0$
hyperfine transition \citep{muk03}, which is detuned by about 1.4~GHz
from the $\Delta F=+1$ transition. The stirring laser must have similar
spectral characteristics as the 2\textsuperscript{nd}-stage cooling
laser and it needs to follow the frequency modulation of the cooling
laser that is applied during the initial phase of the 2\textsuperscript{nd}-stage
MOT operation. 

The two compact 2\textsuperscript{nd}-stage cooling laser pulse distribution
breadboards are based on designs for the stationary Sr lattice clock
at PTB, and use half-inch optical components. To keep the optical
setup simple without requiring a second reference cavity, the stirring
laser is phase-locked to the cooling laser using a beat note between
both. The frequency of this beat note is at 1.4~GHz and is detected
with an avalanche photodiode, integrated on the stirring laser board.
The beat frequency is mixed with a reference frequency from the sixth
harmonic of a 243~MHz DDS. A phase/frequency detector creates an
error signal which is fed back to the piezoelectric transducer (PZT)
(slow part) and the laser diode current (fast part, small corrections)
of the stirring laser. The parameters of the phase-lock were optimized
to obtain a unity-gain bandwidth of 500~kHz. In the setup, commercial
lasers (TOPTICA DL-pro), with long extended cavities for optimized
free-running linewidth are integrated. To save space, the laser heads
were removed from their housings and directly attached to the breadboards.
For the cooling laser, the available laser power is boosted by an
injection-locked laser diode. The output beams are controlled in amplitude
via acousto-optical modulators (AOM), additional mechanical shutters
ensure complete shut-off of the beams, as required to avoid any ac-Stark
shift from stray light during the interrogation of the clock transition.

\begin{figure}
\centering{}\includegraphics[width=0.45\columnwidth]{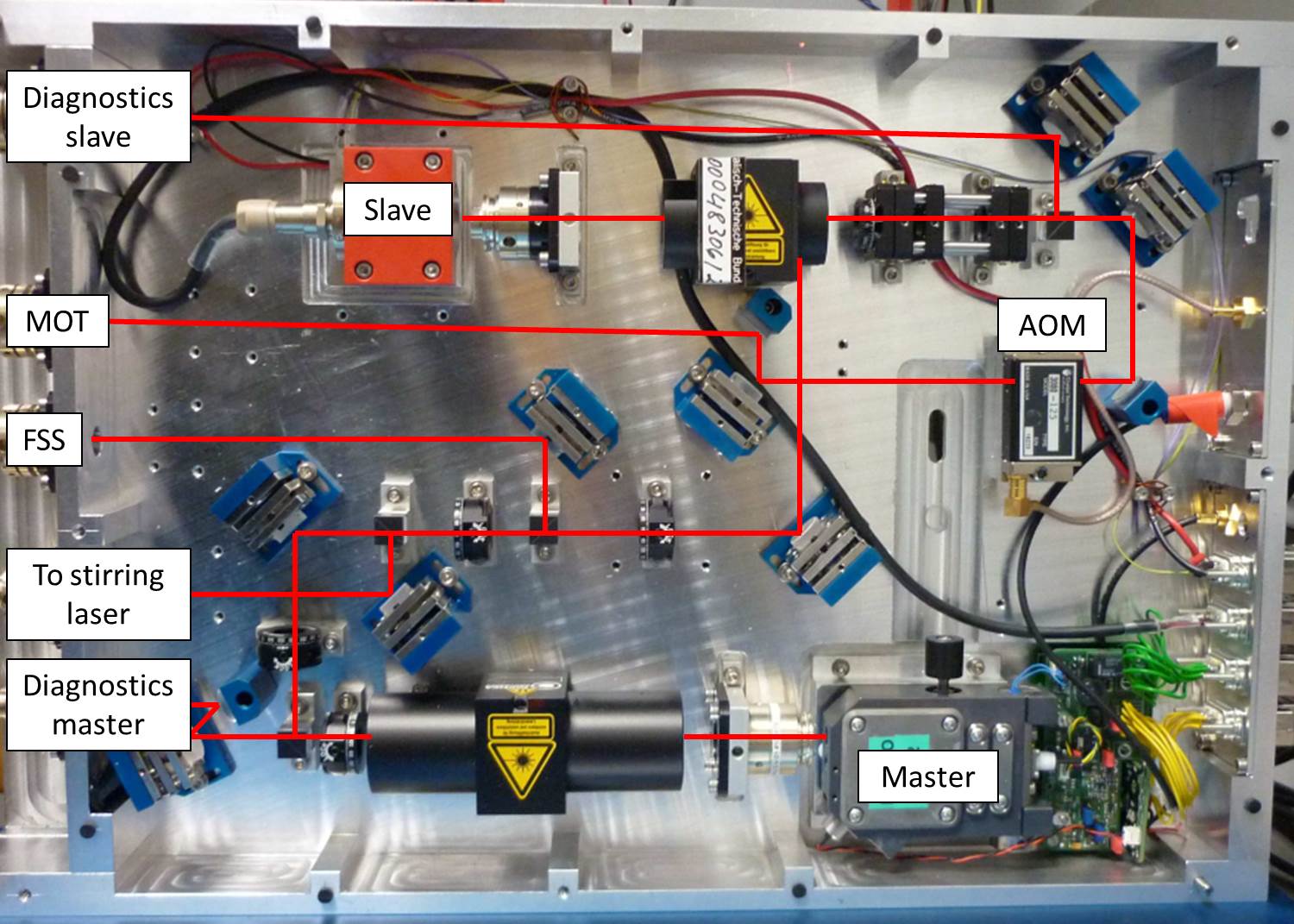}~~~~~~~~~~~~~~~~~~~\includegraphics[width=0.45\columnwidth]{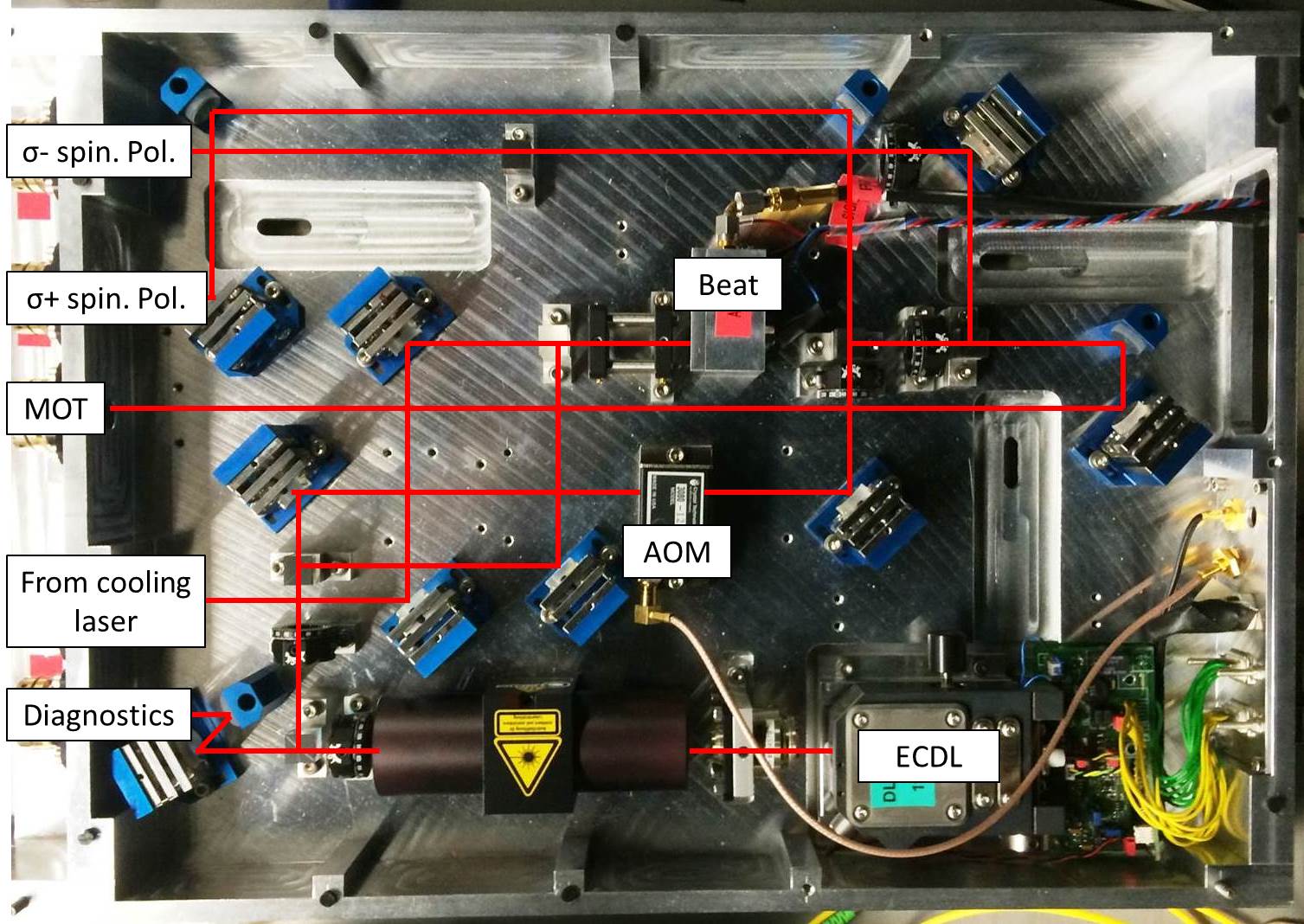}\protect\caption{SHOULD WE CHOOSE DIAGRAMS INSTEAD OF PHOTOS? (color online) Left:
design of the 2\protect\textsuperscript{nd}-stage cooling laser breadboard
(689~nm). Right: design of the stirring laser breadboard (689~nm).\label{fig:689nmLaser}}
\end{figure}

The two 689~nm laser systems comprise optical boards of size 30~cm
\texttimes{} 45~cm \texttimes{} 12~cm and mass 12~kg each (Fig.~\ref{fig:689nmLaser}).
The boards possess fiber outputs into polarization preserving fibers
to the atomics unit and which can provide up to 10~mW each. Additionally,
there is one fiber between both boards, one from the cooling board
to the FSS (1~mW) and two from the stirring board to the atomics
unit for spin-polarization of \textsuperscript{87}Sr. 

\begin{figure}
\centering{}\includegraphics[width=0.6\columnwidth]{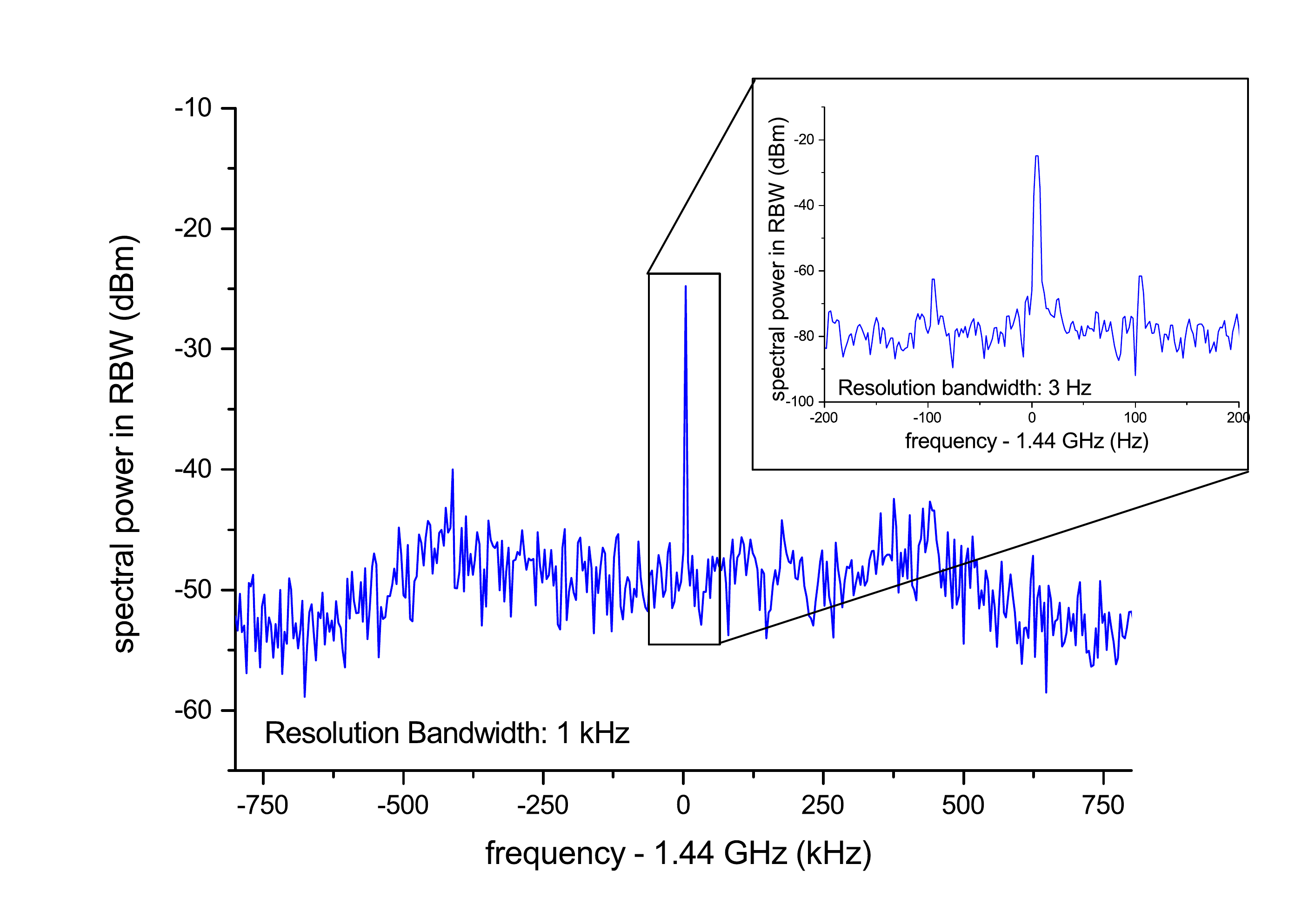}\protect\caption{In-loop beat note of stirring laser phase-locked to the 2\protect\textsuperscript{nd}-stage
cooling laser.\label{fig:beatPLL}}
\end{figure}

The performance of the phase-locked loop (PLL) is indicated in Fig.~\ref{fig:beatPLL}.
As expected for a phase lock with less than one radian phase excursions,
the beat spectrum collapses to a narrow $\delta$-line. The remaining
phase fluctuations show up as a small pedestal, with peaks at the
unity-gain bandwidth of 500~kHz. 

The operation of the clock also requires additional repumping lasers
that, during MOT operation, repump atoms that have decayed from the
$^{1}\mathrm{P_{1}}$ level via the $^{1}\mathrm{D_{2}}$ state to
the metastable $^{3}\mathrm{P_{2}}$ state. The repumpers operate
on the 707~nm transition to the $^{3}\mathrm{S_{1}}$ state, from
where atoms decay to all fine-structure levels of the $^{3}\mathrm{P}$
multiplet. A second laser at 679~nm repumps atom from the $^{3}\mathrm{P_{0}}$
state back to the $^{3}\mathrm{S_{1}}$, so that eventually all atoms
decay back to the ground state via the $^{3}\mathrm{P_{1}}$ state.
The same repumping scheme is used after the clock interrogation, to
pump atoms from the upper clock state back to the ground state for
efficient detection. In the fermionic isotope, both the $^{3}\mathrm{P_{2}}$
and the $^{3}\mathrm{S_{1}}$ level are split by about 5~GHz due
to hyperfine structure \citep{let07a}. Thus, to avoid optical pumping
into dark states, both repumpers are frequency-modulated by a peak-to-peak
amplitude of more than 5~GHz at a frequency of 5~kHz to completely
cover the full hyperfine structure, using a modulation of the cavity
length by PZTs.

\begin{figure}
\includegraphics[scale=0.3]{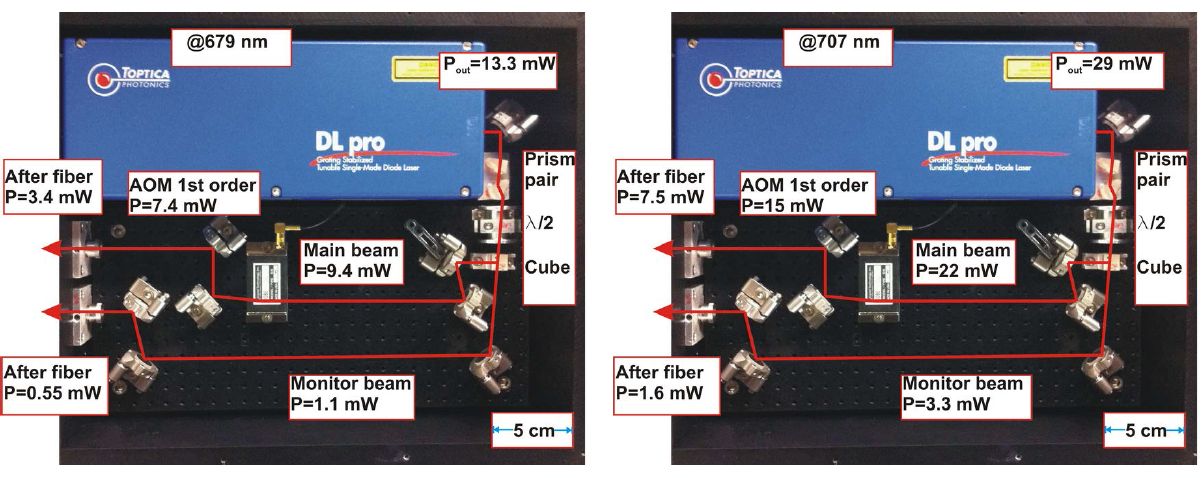}\protect\caption{Repumper breadboards containing the 679~nm laser (left) and the 707~nm
laser (right).\label{fig:Repumper}}
\end{figure}

The repumper breadboards of size 26.5~cm \texttimes{} 32~cm \texttimes{}
12.1~cm, and mass 15~kg (Fig.~\ref{fig:Repumper}) are based on
TOPTICA DL-pro diode lasers that are coupled in two fibers each, one
going to the atomics unit and the other going to a wavemeter for long-term
frequency control. Currently, the lasers are turned on/off using shutters
rather than AOMs, in order to avoid the power loss of AOMs.

\subsection{Laser for the optical dipole trap}

\begin{figure}
\includegraphics[scale=0.2]{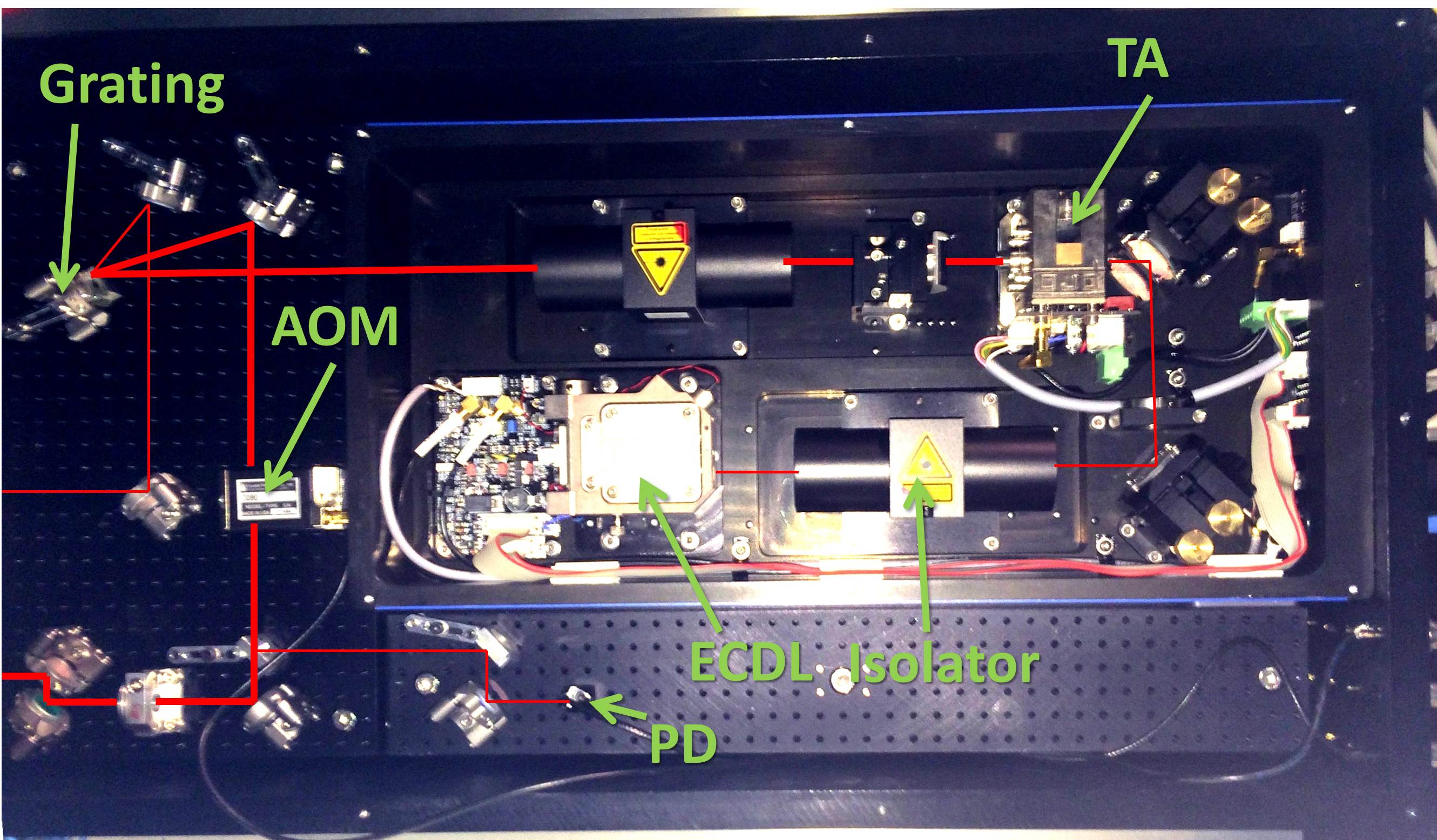}

\protect\caption{The lattice laser (813~nm) and its breadboard. The 813 nm seed laser
(ECDL), is amplified using a tapered amplifier (TA). Its output is
reflected from a diffraction grating acting as a frequency filter.
PD: photodiode.\label{fig:Lattice-laser-breadboard}}
\end{figure}

For the 1D optical dipole trap a high-power laser is required. The
laser (TOPTICA) is a master-oscillator-power amplifier system that
delivers a single-frequency output of 3~W at 813~nm. The laser and
its breadboard (Fig.~\ref{fig:Lattice-laser-breadboard}) have dimensions
50~cm \texttimes{} 20~cm \texttimes{} 20~cm and mass 30~kg. The
breadboard includes a diffraction grating that filters out spurious
spectral background, which is generated by amplified spontaneous emission,
before the light is coupled into a fiber output of the breadboard.

The lattice light is transported via fiber to the atomics unit. There,
it is focused using a telescope designed using commercially available
parts (Thorlabs). The chosen optics yield a focus of the laser beam
with a radius of around 120~\textmu m, centered on the atoms. The
optics can easily be adapted to generate smaller foci. The design
allows overlapping the lattice laser beam with the clock interrogation
beam by having the lattice beam retro-reflected, and the clock laser
introduced through the back side of the retro-reflection mirror. This
arrangement (Fig.~\ref{fig:Vacuum-assembly-for}) further reduces
the size of the apparatus. For 1.5~W of 813~nm light out of the
fiber at the atomics unit and a lattice waist radius of 120~\textmu m
at the position of atoms in the atomic chamber, a lattice trap depth
of 100~$E_{R}$ results, where $E_{R}$ is the lattice recoil energy.

\subsection{Frequency stabilization system (FSS)\label{sub:FSS}}

\begin{figure}[h]
\includegraphics[scale=0.24]{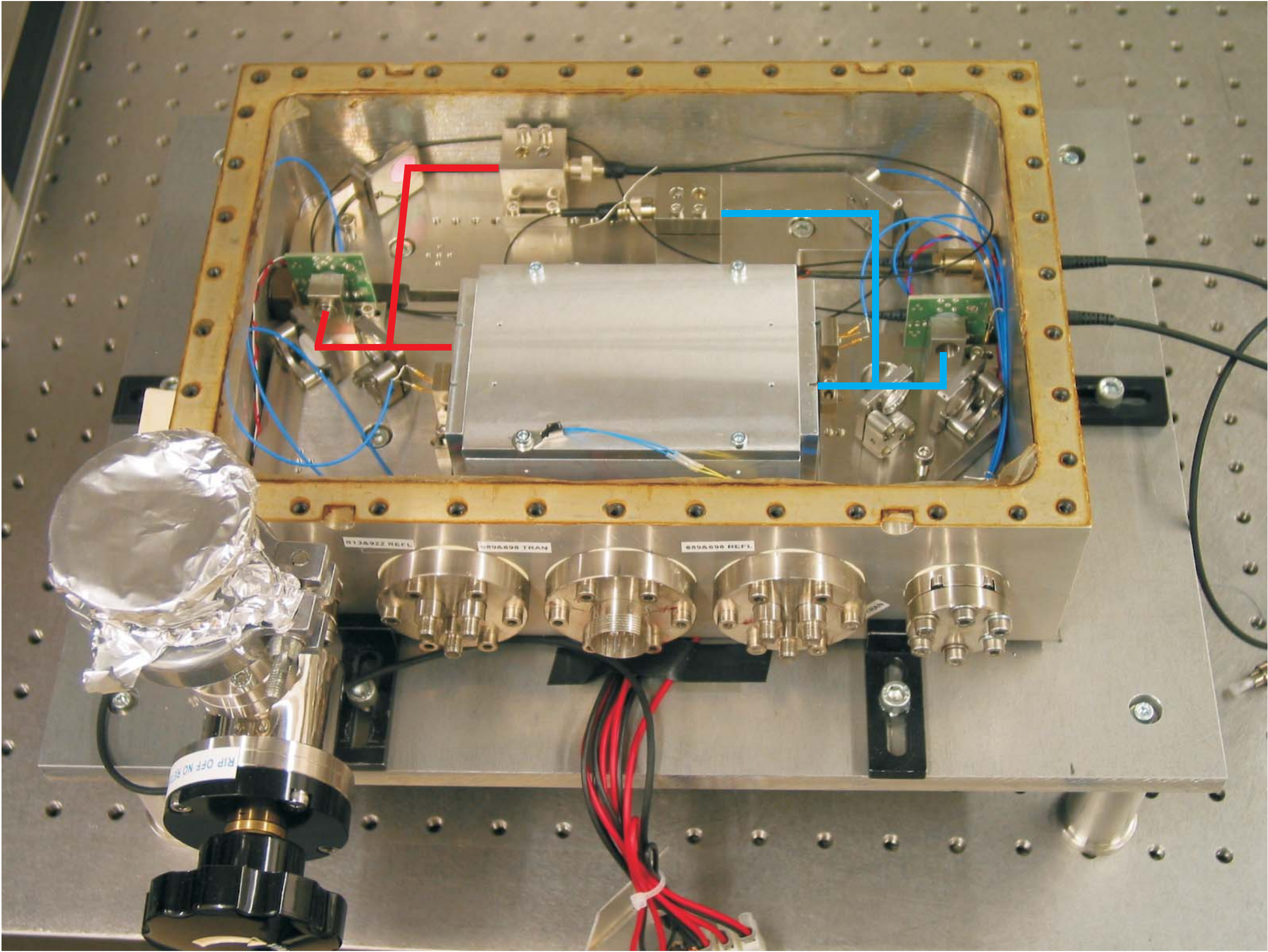}~~~~~~~~~~~~\includegraphics[scale=0.21]{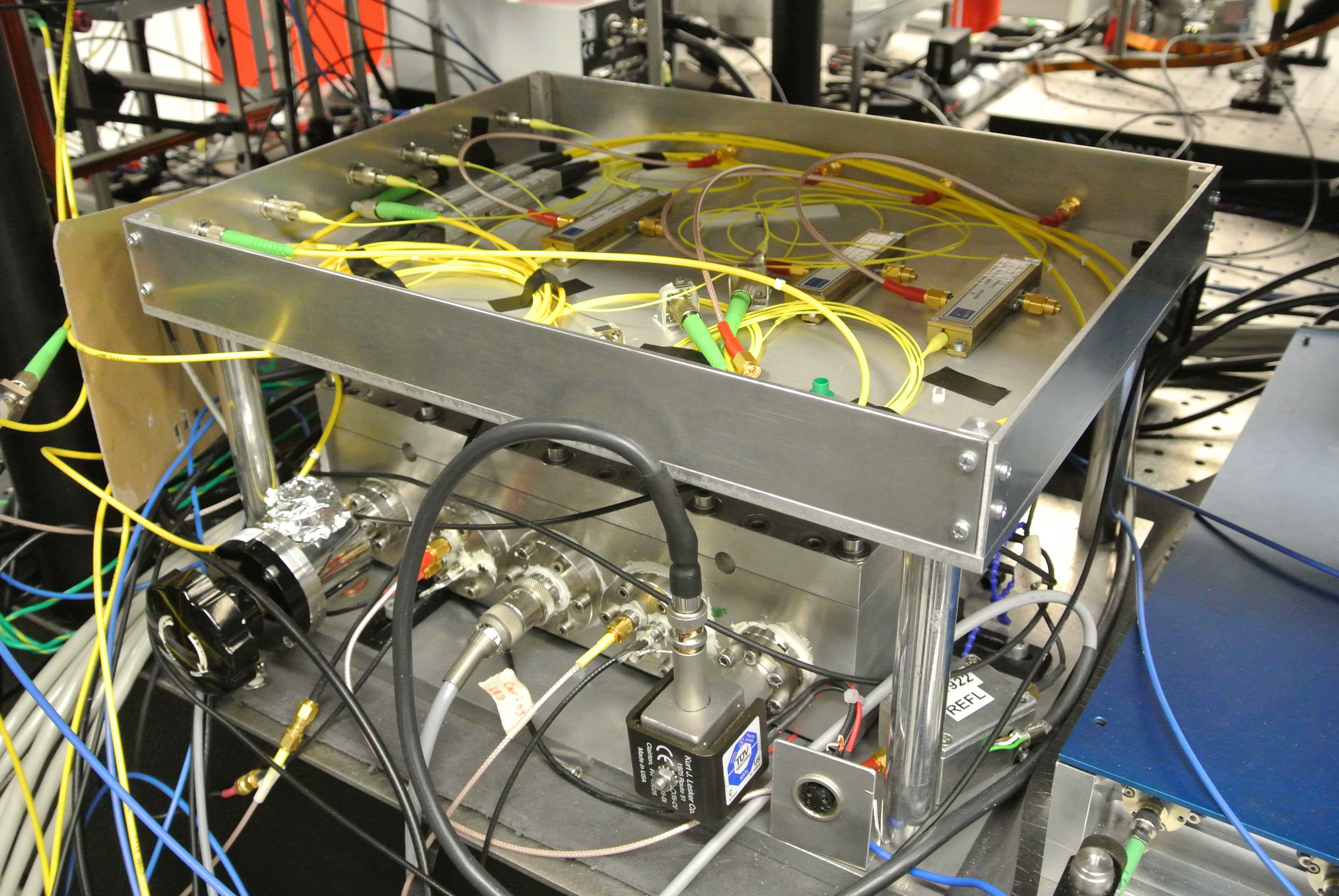}
\protect\caption{The FSS. Left: open vacuum chamber, showing beam paths from the fiber
outcouplers to the cavities and to the reflection error signal photodetectors.
Right: FSS (bottom) and the box containing the fiber phase modulators
(top). \label{fig:The-frequency-stabilization}}
\end{figure}

The 1\textsuperscript{st}-stage cooling laser, the 2\textsuperscript{nd}-stage
cooling laser, and the lattice laser require frequency instabilities
of 1~MHz, 1~kHz, and 10~MHz, respectively, maintained over sufficiently
long time intervals with only an occasional correction from the atomic
signal. A dedicated, compact frequency stabilization system (FSS)
has been developed for this purpose \citep{nev14}. It is based on
a monolithic ultra-low expansion (ULE) block containing three 10~cm
long cavities: one for 922~nm/813~nm, one for 689/698~nm, and one
for the 679~nm/707~nm repumpers. The latter is unused at present,
since the passive frequency stability of the repumpers is sufficient.
 The ULE block is embedded in a 30~cm \texttimes{} 20~cm \texttimes{}
10~cm vacuum chamber which also contains optics and photodetectors.
The laser waves are input into the chamber via fiber feed-throughs.
Figure~\ref{fig:The-frequency-stabilization} shows the system. Together
with a box containing the waveguide phase modulators and fiber connectors,
the total volume occupied is 50~cm \texttimes{} 30~cm \texttimes{}
25~cm. The total mass is approx. 25~kg.

The cavities are not tunable. In order for the lasers to be tunable
to their respective atomic transitions, an offset-locking technique
was implemented. Each laser wave is phase-modulated at a frequency
produced by a DDS. One of the two sidebands is locked to the respective
reference cavity. By computer-controlled tuning of the DDS frequency,
the optical frequency of the carrier wave can thus be tuned without
loss of lock. In order to obtain robust PDH (Pound-Drever-Hall) error
signals for frequency lock, an additional phase modulation of the
DDS frequency is used.

The 922~nm and 813~nm lasers are locked using the DIGILOCK 110 (TOPTICA).
The 689~nm laser is locked using a home-made proportional-integral-differential
(PID) servo electronics. The measured linewidths, less than 1~MHz
for the 922~nm and 813~nm lasers and less than 1~kHz (1~min time
scale) for the 689~nm laser, are narrower than the required ones
\citep{nev14}. The typical long-term drift of the 689~nm frequency
is on the order of 0.5~Hz/s, since the ULE block is not operated
at the temperature of vanishing thermal expansion coefficient. This
drift value will eventually shift the laser out of the 7~kHz-wide
atomic resonance. We observed that the typical frequency drift in
day-to-day operation is of the order of $10-20$ kHz (on the 689~nm
cavity). It can be corrected using information from the Sr atomic
signal. Another possibility is to interrogate the 689~nm cavity by
a sideband of the independently stabilized clock laser (698~nm) and
to use the value of its sideband frequency to correct the frequency
of the 689~nm sideband DDS. The FSS is very robust: it can be tilted
while the frequency locks are maintained.

\subsection{Clock laser}

\begin{figure}
\includegraphics[width=0.6\columnwidth]{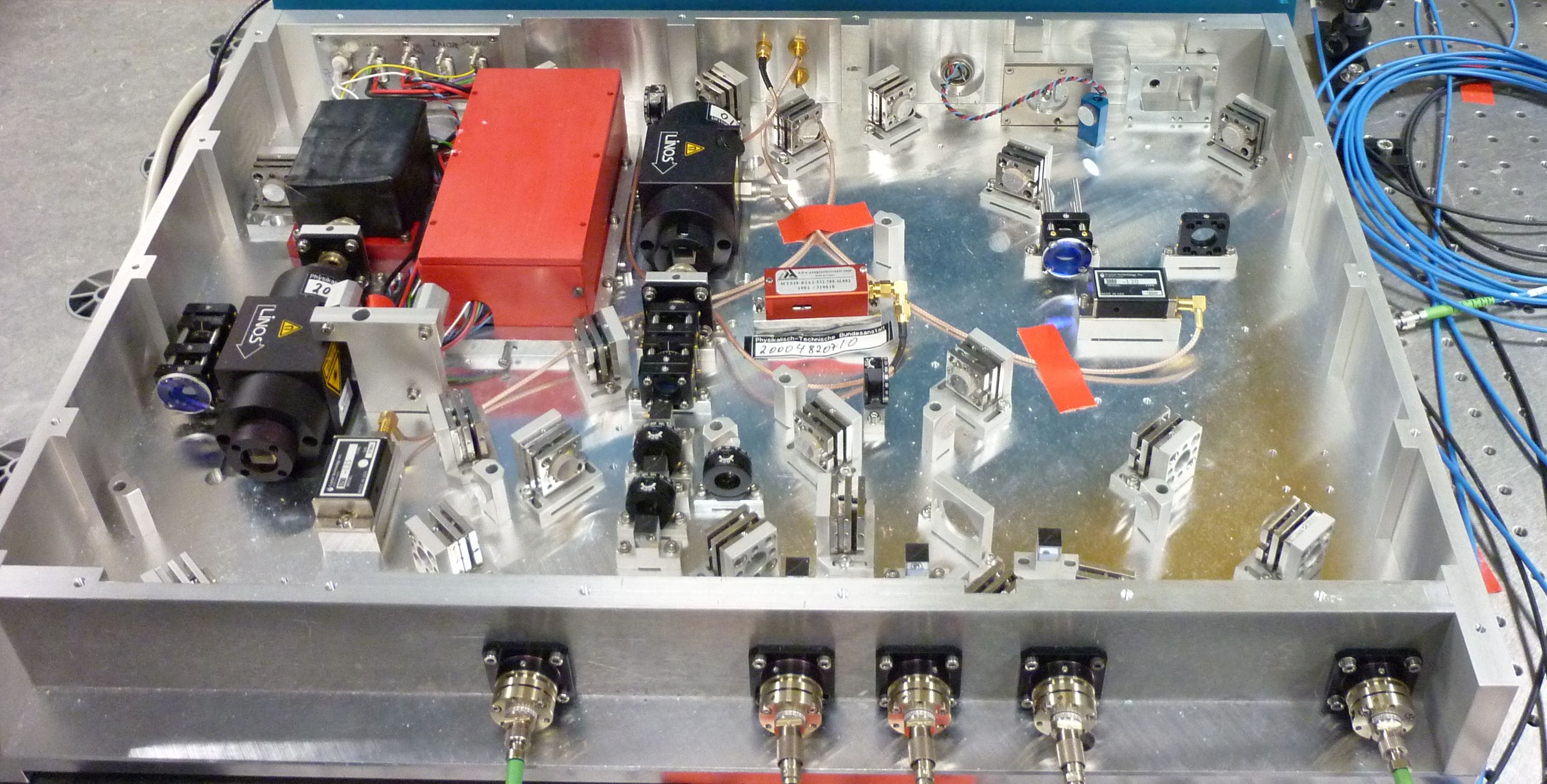}

\protect\caption{The 698~nm laser breadboard. \label{fig:698 nm laser breadboard}}
\end{figure}

The 698~nm laser for interrogation of the atomic clock transition
consists of the laser breadboard, the reference cavity, and the electronics
modules. 

The laser breadboard of size 60~cm \texttimes{} 45~cm \texttimes{}
12~cm and mass 20~kg (Fig.~\ref{fig:698 nm laser breadboard})
provides fiber-coupled outputs for diagnostics by a wavemeter or for
the FSS, for the frequency comb, and switchable light with adjustable
power for exciting the atoms. The outputs for the comb and for the
atomics unit contain the necessary optics for a cancellation of the
fiber noise by detecting the round-trip phase through the fiber, which
is possible even in pulsed mode during the interrogation of the clock
transition \citep{fal12}. The clock laser breadboard initially contained
a low-power filter-stabilized extended-cavity diode laser \citep{bai06}.
A slave laser was injection-locked to its output, boosting the power
to 10~mW. During the course of the SOC2 project, more powerful filter-stabilized
698-nm lasers were developed at LUH, that do not require the additional
slave. 

The clock laser is tightly frequency-locked to an ultrastable Fabry-Perot
cavity via a PDH feedback loop, in order to narrow down the spectrum.
This leads to a long coherence time of the laser wave, thus enabling
long interrogation times of the atoms and therefore Fourier-limited
atomic resonances, with widths potentially on the order of 1~Hz.
To achieve a narrow laser linewidth, the cavity must exhibit a high
finesse, a low sensitivity to vibrations and must be under vacuum
in order to avoid fluctuations of its optical length. The temperature
stabilization of the cavity must be sufficiently accurate to reduce
the frequency drift during the spectroscopy phase to a level such
that locking errors are small compared to the targeted accuracy. Once
referenced to such a cavity, the short-term fractional frequency stability
of the laser is typically in the $\ensuremath{10^{-15}}$ range. The
long-term fluctuations (time scale larger than a few seconds) of the
laser frequency are corrected via a slow lock to the atomic clock
transition, while the short-term laser fluctuations still contribute
to the clock instability through the Dick effect \citep{dic87}.

\begin{figure}
\includegraphics[bb=0bp 0bp 320bp 230bp,clip,width=0.5\columnwidth]{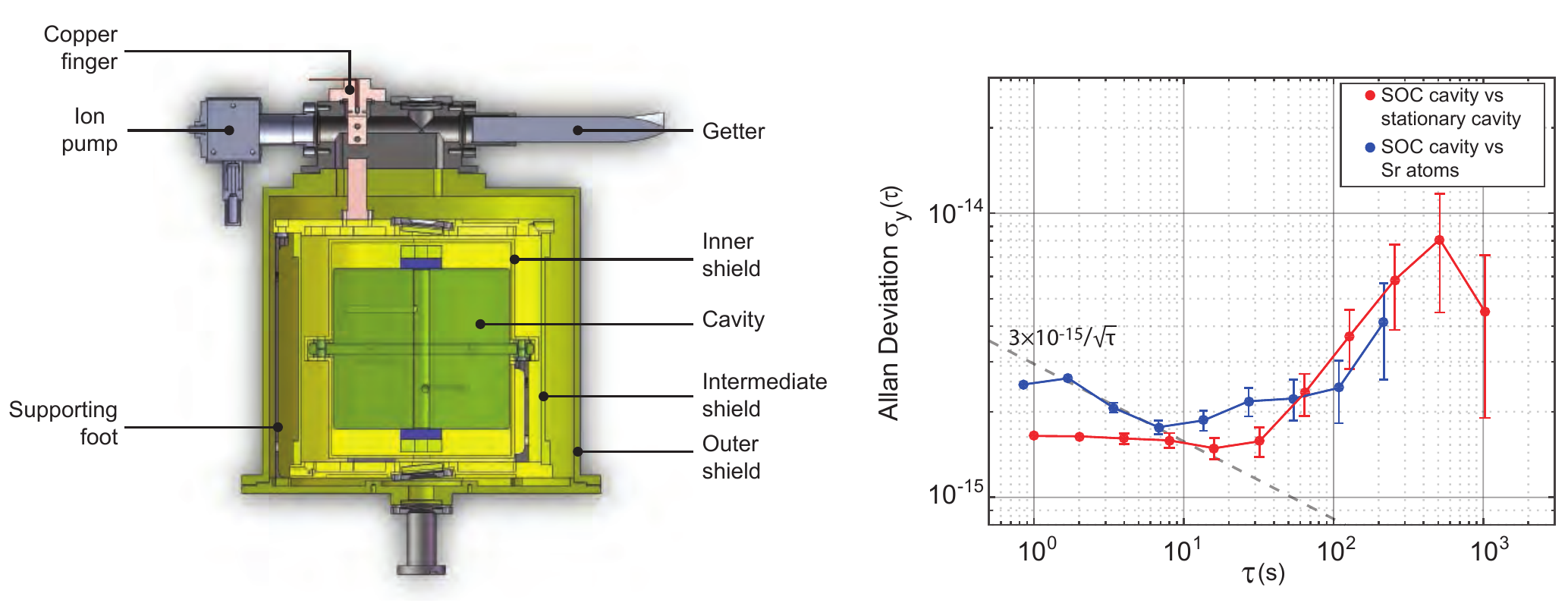}\includegraphics[width=0.32\columnwidth]{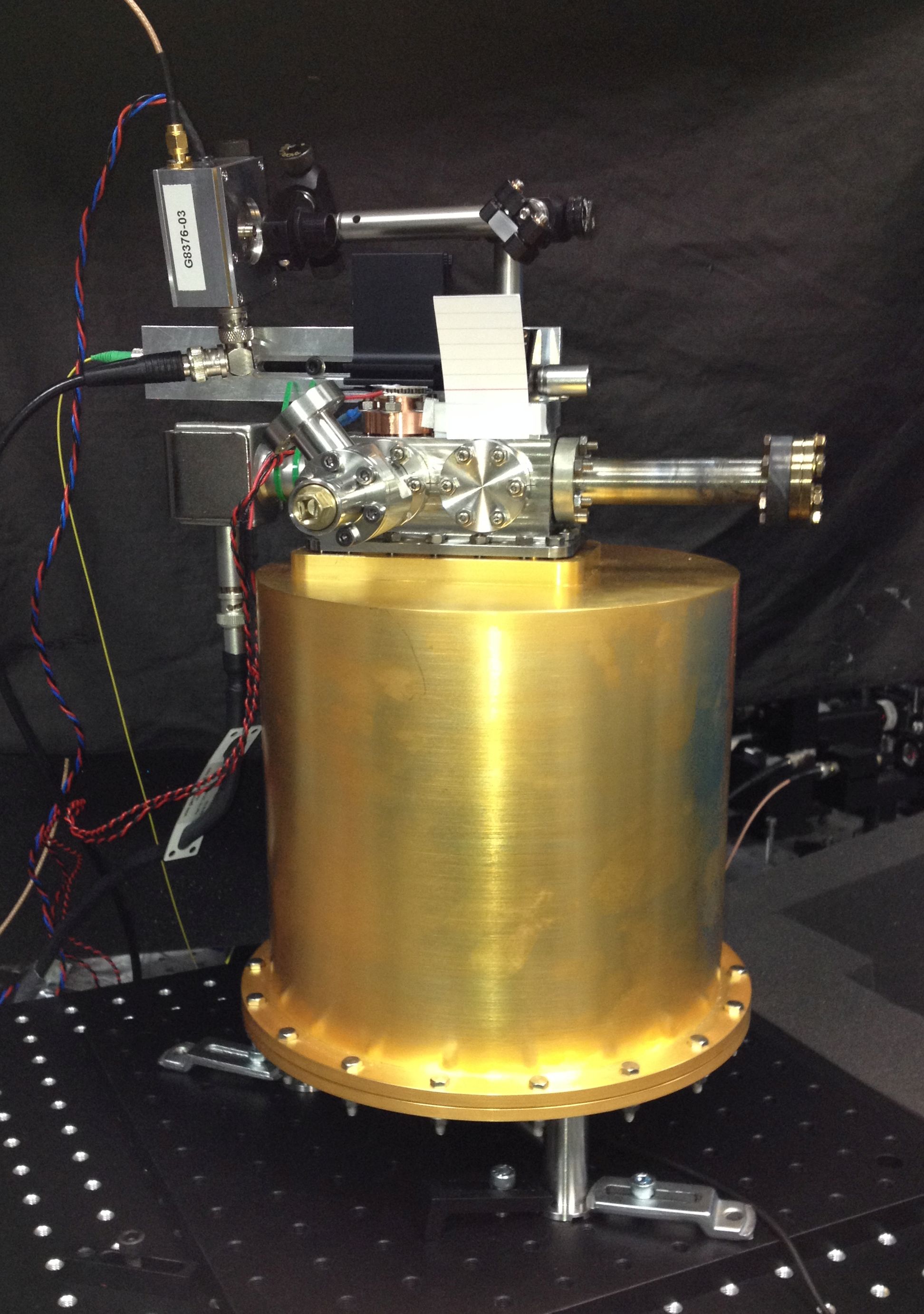}

\protect\caption{Left: Section view of the cavity and its vacuum environment. Right:
the clock laser cavity assembly. \label{fig:assembly}}
\end{figure}

A compact ultrastable cavity setup was developed with consideration
of robustness against mechanical disturbances and against uncontrolled
temperature changes. The 10~cm long reference cavity is made of a
ULE glass spacer and uses optically contacted fused silica mirror
substrates in order to reduce the thermal noise floor. A ULE glass
ring is contacted to the back of each mirror substrate in order to
compensate for the differential temperature sensitivity between the
spacer and the mirrors \citep{leg10}. The cavity is mounted vertically,
tightly fixed to three gold-coated aluminum shields \citep{argence2012prototype}
as shown in Fig.~\ref{fig:assembly}. The outer shield is vacuum-tight,
and includes a copper finger that controls the temperature of the
intermediate shield. The shields are connected one to another by three
titanium feet, therefore ensuring a very low heat exchange by conduction
between the shields \citep{thompson2011flight}. The overall mass
of the vacuum system plus cavity is 9~kg, and the volume is 5~liter.
The transport of the setup by car over 800~km did not result in any
significant misalignment of the optics for coupling light into the
cavity.

Integration

\begin{figure}
\includegraphics[scale=0.5]{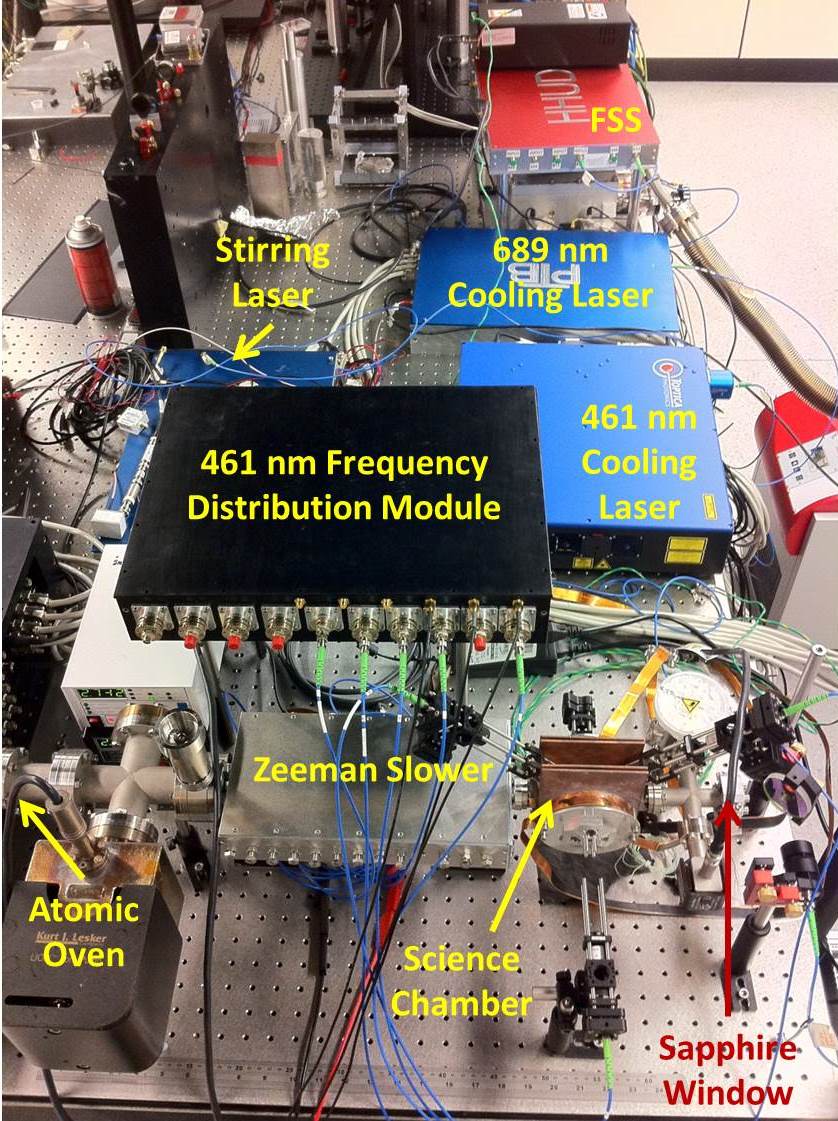}

\protect\caption{\label{fig:Atomics-unit-integrated}Atomics unit integrated with the
laser systems, except for the clock laser system. }
\end{figure}

The laser breadboards and the atomics unit are currently collocated
on one optical table, see Fig.~\ref{fig:Atomics-unit-integrated}.
At the atomics unit all the necessary laser light is brought in by
fibers and connected to collimation packages. For the 3D MOT, dichroic
couplers including quarter-wave plates suitable for cooling with 461~nm
and 689~nm radiation collimate the overlapped 1\textsuperscript{st}-stage
and 2\textsuperscript{nd}-stage trapping/cooling beams. These telescopes
are attached to the chamber using custom-built, highly versatile yet
stable adapters. The telescopes, retro-reflection couplers and adapters
all have low sensitivity to misalignment and to vibrations. For the
slower, repumper and detection beams, similar couplers are also employed.

\section{Preliminary Characterization Results}

\subsection{1\protect\textsuperscript{st}-stage MOT}

The 461~nm distribution module delivers, from three fibers, 3~mW
each to three retroreflected MOT beams. The output of each fiber is
collimated to a diameter of 10~mm. A fourth fiber delivers 65~mW
for the Zeeman slowing beam, which is slightly focused onto the atomic
source in order to improve the slowing and has a diameter upon entrance
to the chamber of 12 mm. 

\begin{figure}
\includegraphics[width=0.4\columnwidth]{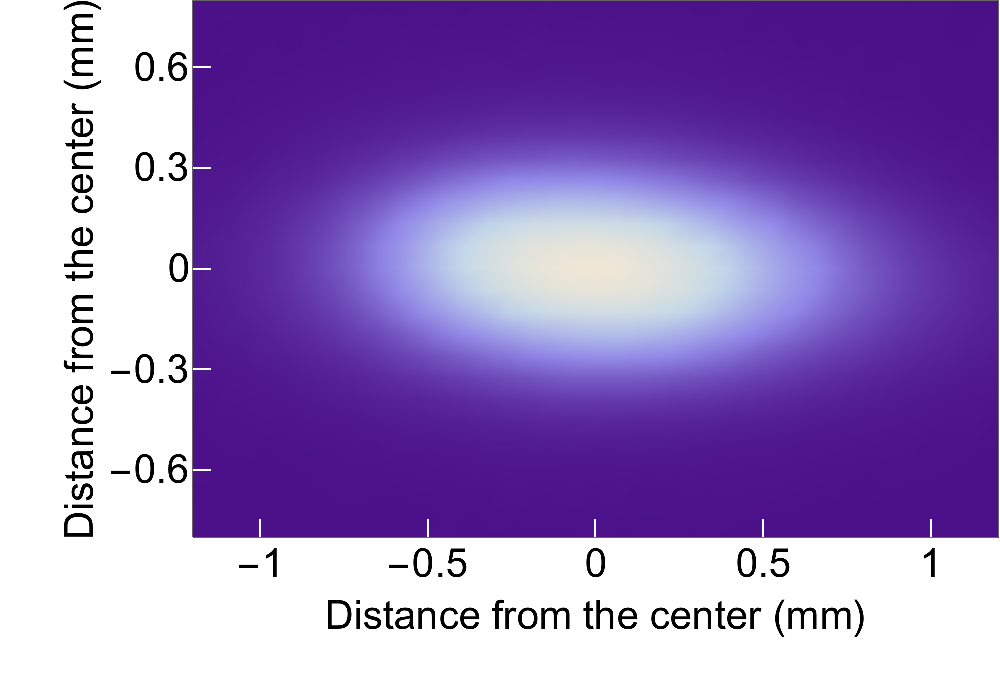}~~~~~~~~~~~~~~~~~~~~~\includegraphics[width=0.4\columnwidth]{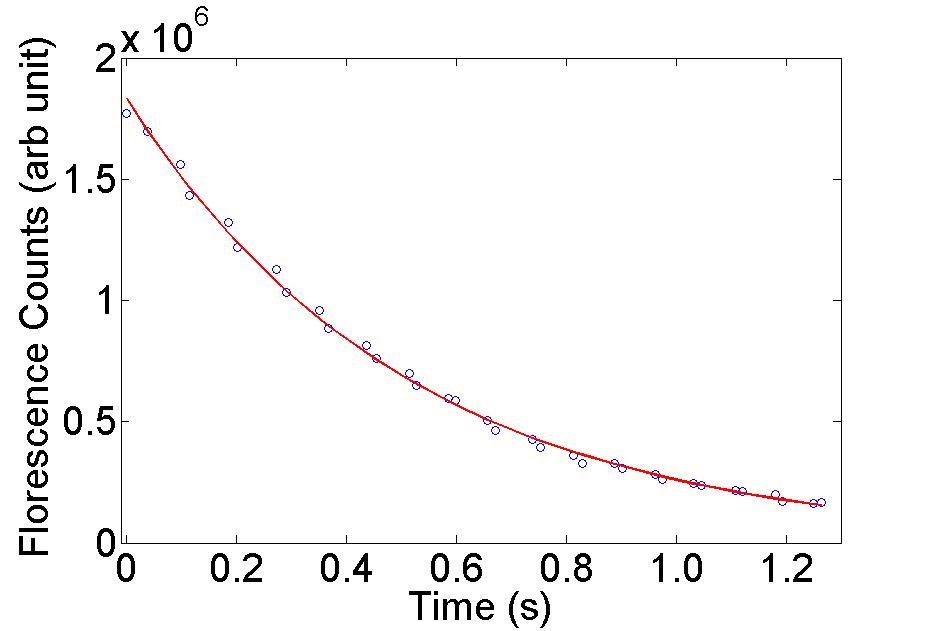}

\protect\caption{\label{fig:BlueMOT}Left: fluorescence image of the 1\protect\textsuperscript{st}-stage
MOT containing $8\times10^{6}$ atoms. Size is approximately 1~mm
in diameter. The temperature of the MOT is measured to be $1.2\pm0.1$~mK.
Right: lifetime measurement of the atoms in the MOT, yielding 530~ms.
}
\end{figure}

The AOMs in the distribution unit for the MOT and Zeeman slowing beams
can be controlled via computer and thus the detuning and intensity
of the respective beams can be optimized to further cool and compress
the 1\textsuperscript{st}-stage MOT. The magnetic field gradient
produced by the Helmholtz coils is also computer-controlled. After
optimization, the 1\textsuperscript{st}-stage MOT provides approx.
$8\times10^{6}$ $^{88}$Sr atoms and is approx. 1~mm in diameter.
The temperature obtained after an optimized cooling sequence, which
involved ramping down the intensity and detuning of the MOT beams
along with the magnetic field gradient, was determined to be $1.2\pm0.1$~mK.
The measured lifetime is approximately 530~ms (Fig.~\ref{fig:BlueMOT}).
The blue MOT consistently runs for at least 12 hours without interruption.

\subsection{2\protect\textsuperscript{nd}-stage MOT}

The laser light from the 689~nm laser is divided into three waves
by a 1-to-3 fiber splitter (Evanescent Optics). This ensures  low
overall loss and a constant power balance between the beams. The three
output fibers are overlapped with the blue MOT beams in the dichroic
telescopes on the MOT chamber. This ensures that the alignment of
the red beams will be optimized along with that of the blue beams.
Each red MOT beam is 3.3~mW in power and 10~mm in diameter. Compared
to the 1\textsuperscript{st}-stage MOT, the magnetic field gradient
for the 2\textsuperscript{nd}-stage MOT is only 0.65~mT/cm. 

\begin{figure}
\includegraphics[width=0.3\columnwidth]{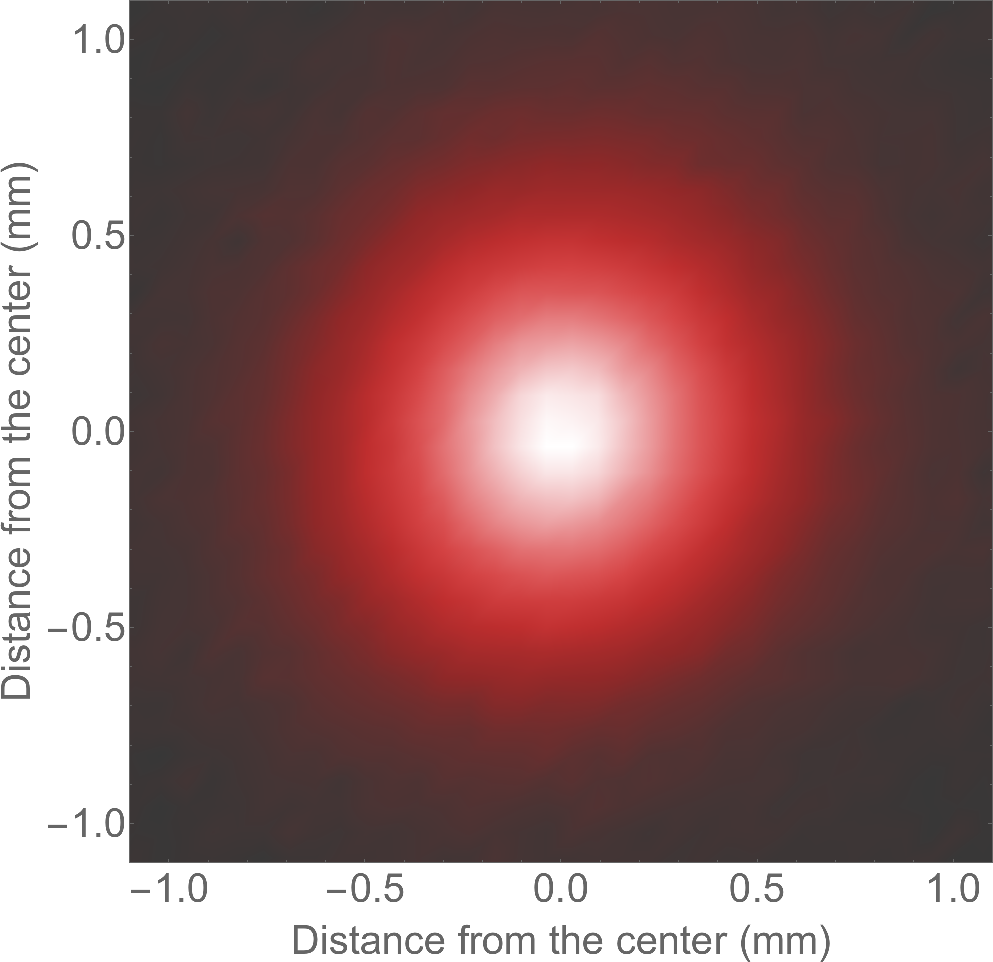}~~~~~~~\includegraphics[bb=45bp 58bp 1000bp 965bp,clip,width=0.3\columnwidth]{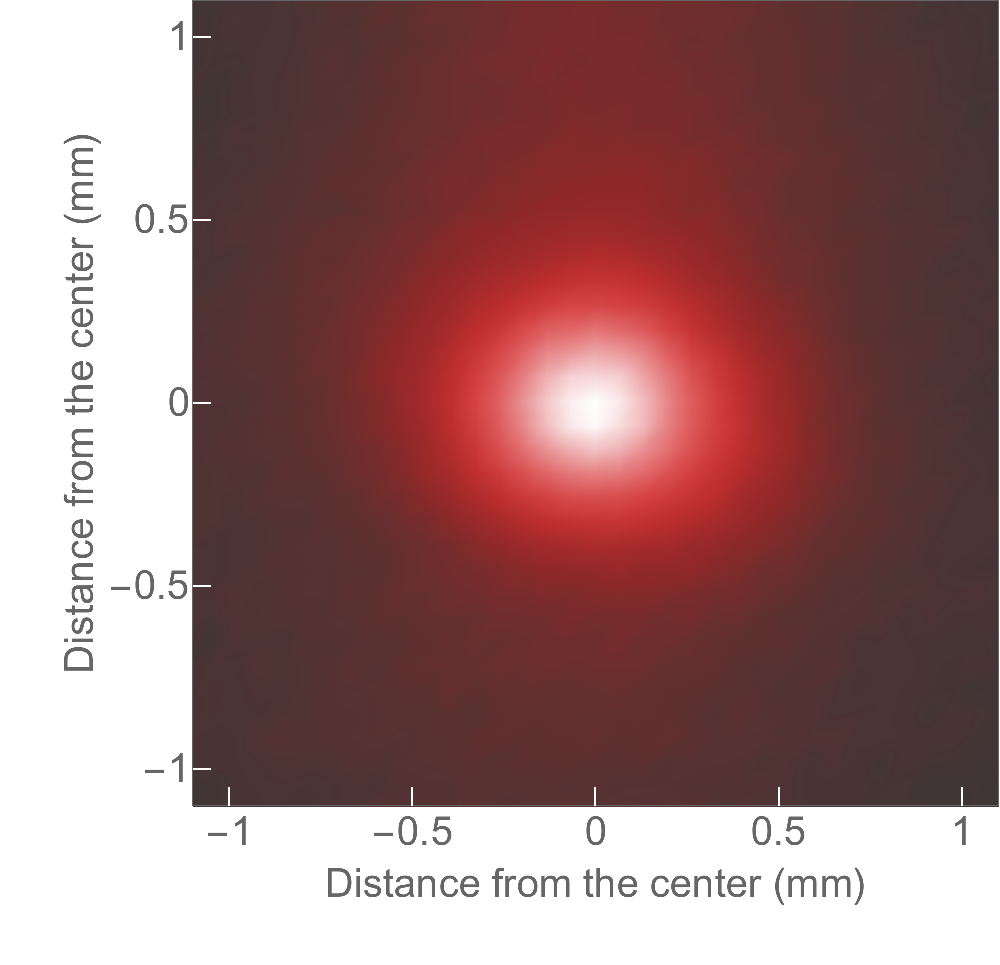}

\protect\caption{\label{fig:RedMOT}Fluorescence images of the 2\protect\textsuperscript{nd}-stage
MOT. Left: broad-band phase, $2.3\times10^{6}$ atoms; size is approximately
0.7~mm in diameter. Temperature is $170\pm10\mu$K. Right:
single-frequency phase, containing $1.8\times10^{6}$ atoms at a temperature
below 2~$\mu$K.}
\end{figure}
As mentioned above, in the initial, broad-band, 2\textsuperscript{nd}-stage
cooling phase, the 689~nm radiation must be spectrally broadened
in order to overlap the Doppler spectral width of the 1.2~mK ``warm''
atoms, allowing capturing a sufficient fraction of the atoms into
the 2\textsuperscript{nd}-stage MOT. The double phase modulation
in the FSS is used for this purpose. The 689~nm laser is locked with
a sideband to the FSS cavity. The sideband offset frequency is typically
around 80 MHz. The carrier is detuned by 2.5~MHz from the cooling
transition. The sideband offset frequency is modulated with a peak-peak
amplitude of 5~MHz at a frequency of 30~kHz in order to achieve
the required amount of spectral broadening. With 30~ms of broad-band
phase, we obtained a transfer efficiency of at least 40\% from the
1\textsuperscript{st}-stage to the 2\textsuperscript{nd}-stage MOT
, see Fig.~\ref{fig:RedMOT} (left).

The following single-frequency phase lasts for another 30 ms, the
689~nm laser now being detuned by 600~kHz but not modulated, and
with each MOT beam attenuated to approximately 100~$\mu$W. We obtain
80\% transfer efficiency from broadband to single-frequency MOT, see
Fig.~\ref{fig:RedMOT} (right). This gives an overall transfer efficiency
from 1\textsuperscript{st}-stage MOT to single-frequency 2\textsuperscript{nd}-stage
MOT of at least 30\%.

\subsection{Clock laser}

\begin{figure}
\includegraphics[bb=327bp 8bp 595bp 230bp,clip,width=0.45\columnwidth]{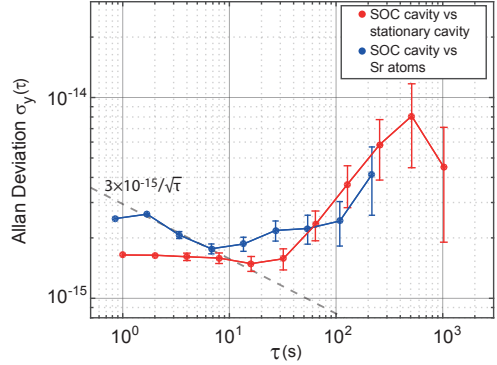}

\protect\caption{Combined fractional frequency instability of the SOC cavity versus
a stationary reference cavity (red) and versus the clock transition
of lattice-trapped Sr atoms (blue). Both sets of data were corrected
for a second-order drift. \label{fig:clock_assembly-1}}
\end{figure}

The clock laser breadboard (without the cavity described above) was
successfully tested with an earlier reference resonator \citep{vog11}
for the characterization of the original SOC physics package \citep{pol14}.
With this system the clock transition in \textsuperscript{88}Sr could
be observed with 10~Hz linewidth in a different experiment.

In order to evaluate the stability of the new cavity, we have locked
a (different, stationary) master laser at 698~nm to a stationary
ultrastable cavity, with a known flicker noise level of $\ensuremath{6\times10^{-16}}$,
and offset-locked the resulting light to the new cavity. The frequency
stability of the offset reflects the combined stability of the two
cavities (red line, Fig.~\ref{fig:clock_assembly-1}). Overall, the
flicker floor is $\ensuremath{1.7\times10^{-15}}$, corresponding
to $\ensuremath{1.6\times10^{-15}}$ fractional instability for the
new cavity. This value is still above the expected thermal noise level
of $\ensuremath{6\times10^{-16}}$. Residual pressure fluctuations
likely lead to an optical path length jitter of this magnitude in
the cavity. This could be improved further with a larger pumping capacity.
The setup is temperature-controlled close to room temperature ($\ensuremath{21~^{\circ}\mathrm{C}}$),
and the maximum drift is less than 80~mHz/s, which can easily be
compensated. The $\ensuremath{1}$~K peak-peak daily variation of
the laboratory temperature leads to a 500~Hz frequency change, corresponding
to a sensitivity around $\ensuremath{10^{-12}\,\mathrm{K}^{-1}}$.
With an estimated coefficient of thermal expansion of the cavity of
$\ensuremath{10^{-9}\,\mathrm{K}^{-1}}$ this indicates that the three
thermal shields damp temperature fluctuations by 3 to 4 orders of
magnitude. At SYRTE, the probing of lattice-trapped Sr atoms with
a stationary laser locked to this cavity lead to a stability of $\ensuremath{3\times10^{-15}}$
at 1~s (black line, Fig.~\ref{fig:clock_assembly-1}). This number
is in good agreement with the expected contribution of the Dick effect,
corresponding to a flicker floor of $\ensuremath{1.6\times10^{-15}}$.

\section{Summary and Outlook}

We developed a novel Sr lattice clock apparatus of modular design
and consisting of compact subunits. The subunits are designed for
mechanical robustness. The total mass is approx. 200~kg, excluding
electronics. Although the apparatus is not a prototype of a space
clock, the value of this physical parameter is already in a range
compatible with the intended space application on the ISS. At present,
the apparatus can reliably trap atoms in the 1\textsuperscript{st}-stage
and single-frequency 2\textsuperscript{nd}-stage MOTs. Optimization
of the MOT is ongoing. The next step is achieving trapping in the
optical lattice, after which the apparatus will be transferred to
PTB for further optimization and characterization with respect to
stationary optical clocks. For transportability, all subunits, including
the atomics package, will be installed in a vibration-isolated and
fully transportable rack of less than 970~liter volume (1560~l including
electronics). \medskip{}

\textbf{Acknowledgments:} The research leading to these results has
received initial funding by ESA, DLR and other national sources. Current
funding has been provided by the European Union Seventh Framework
Programme (FP7/2007-2013) under grant agreement n\textdegree{} 263500.
The work at PTB was also funded by the European Metrology Research
Programme (EMRP) under IND14. The EMRP is jointly funded by the EMRP
participating countries within EURAMET and the European Union. J.~H.
and D.~S. acknowledge the funding from the EPSRC (EP/L001713/1) and
Qtea (FP7-People-2012-ITN-Marie-Curie Action ``Initial Training Network
(ITN)''), respectively. S.~O. was funded by the Marie-Curie Action
ITN ``FACT''. S.~V. acknowledges funding from the DFG within the
RTG 1729.

\bibliographystyle{apsrev4-1}
\bibliography{bibfile_final}

\end{document}